\documentclass[a4paper,11pt]{article}
\pdfoutput=1 \usepackage{jheppub_2} 
\usepackage[T1]{fontenc} 
\usepackage[utf8]{inputenc}
\usepackage{color,subfigure}
\usepackage{graphicx, multirow,soul,url,amsmath,amsfonts,amssymb,mathrsfs,amsfonts}
\usepackage[all]{hypcap}
\usepackage{url}
\usepackage{bbm}
\usepackage{cancel}
\usepackage{subfigure}
\usepackage{slashed}
\usepackage[dvipsnames]{xcolor}
\usepackage{multicol, blindtext}
\usepackage[section]{placeins}
\usepackage[version=3]{mhchem}
\usepackage{caption}
\usepackage{lipsum}
\usepackage{hyperref}
\usepackage{float}
\usepackage{mattens}
\usepackage{mathtools}
\usepackage{footnote}
\usepackage{amsmath,amssymb,amsthm,amsxtra,overpic,bbm,bm,epsfig}
\usepackage{color,subfigure}
\usepackage[splitrule]{footmisc}
\usepackage{bbold}
\usepackage{amsmath}

\newcommand{\equaref}[1]{Eq.~(\ref{#1})}

\newcommand{\secref}[1]{Section~\ref{#1}}
\newcommand{\appref}[1]{Appendix~\ref{#1}}
\newcommand{\tabref}[1]{Table~\ref{#1}}

\newcommand{\SU}{\,{\rm SU}}

\preprint{IPPP/18/40 \hspace{25.5mm} FERMILAB-PUB-18-222-T \hspace{25.5mm} UCI-TR-2018-05}

\title{\boldmath Assessing Perturbativity and Vacuum Stability in High-Scale Leptogenesis}

\author[a]{Seyda Ipek,}
\author[b]{Alexis D. Plascencia}
\author[c]{and Jessica Turner}

\affiliation[a]{Department of Physics and Astronomy, University of California, Irvine
4129 Frederick Reines Hall, Irvine, CA 92617-4575, U.S.A.}
\affiliation[b]{Institute for Particle Physics Phenomenology, Department of
Physics, Durham University, South Road, Durham DH1 3LE, United Kingdom.}
\affiliation[c]{Theoretical Physics Department, Fermi National Accelerator Laboratory, P.O. Box 500, Batavia, IL 60510, USA.}

\emailAdd{sipek@uci.edu}
\emailAdd{a.d.plascencia-contreras@durham.ac.uk}
\emailAdd{jturner@fnal.gov}

\abstract{We consider the requirements that all coupling constants remain perturbative and the electroweak vacuum metastable up to the Planck scale in high-scale thermal leptogenesis, in the context of a type-I seesaw mechanism. We find a large region of the model parameter space that satisfies these conditions in combination with producing the baryon asymmetry of the Universe. We demonstrate these conditions require ${\rm Tr}[Y_N^\dagger Y_N] \lesssim 0.66$ on the neutrino Yukawa matrix. We also investigate this scenario in the presence of a large number $N_F$ of coloured Majorana octet fermions in order to make quantum chromodynamics asymptotically safe in the ultraviolet.}

\begin{document} 
\maketitle
\flushbottom

\section{Introduction}
\label{sec:intro}

The Standard Model (SM) cannot be a complete fundamental theory. In the SM the hypercharge coupling increases with energy in the deep ultraviolet (UV) reaching a Landau pole for a finite value of the energy scale. In addition to this ultraviolet shortcoming, there are a number of phenomenological issues such as the baryon asymmetry of the Universe (BAU), dark matter, inflation, strong CP-violation and neutrino masses that further motivate physics beyond the SM. \\

Although there has been impressive progress in understanding neutrino physics in the last decades, the nature of neutrinos -- namely, whether they are Majorana or Dirac particles -- and the origin of their masses remain unknown. One of the most economical methods to generate neutrino masses 
 is via the type-I seesaw mechanism \cite{Minkowski:1977sc, Yanagida:1979as, GellMann:1980vs, Mohapatra:1979ia}. In this mechanism one introduces heavy Majorana neutrinos which are singlets under the SM gauge group.  In addition to providing an explanation for small but non-zero neutrino masses, the type-I seesaw also provides a solution to the BAU, due to a new source of CP-violation present in the out-of-equilibrium decays of the heavy Majorana neutrinos in the early Universe \cite{Fukugita:1986hr}. \\

The key question we address in this work is: can a minimal extension of the SM by three heavy Majorana neutrinos, simultaneously addressing neutrino masses and the observed matter--antimatter asymmetry,  be self-consistent up to the Planck scale without the addition of new degrees of freedom? 
In order to answer this question,  we investigate the self-consistency of high-scale leptogenesis up to the Planck scale by
performing a systematic study of the renormalisation group (RG) of this model to assess whether the coupling constants remain perturbative and the electroweak vacuum metastable. 
In addition, we  ascertain the validity of the model up to $\sim 10^{40}$ GeV,  the scale at which the hypercharge develops a Landau pole. Moreover, we extend this study to consider a scenario where we introduce a large number, $N_F$, of coloured fermions  to 
achieve asymptotic safety in the strong sector of the SM.\\

There have been a number of works which have studied similar constraints in the type-I seesaw  \cite{Casas:1999cd,EliasMiro:2011aa,Chen:2012faa,Rodejohann:2012px,Bambhaniya:2016rbb,Ghosh:2017fmr} 
and including the presence of the QCD axion \cite{Ballesteros:2016xej,Salvio:2015cja}.
In our work, we extend their study by explicitly solving the density matrix equations for thermal leptogenesis and studying the high-energy robustness of the associated parameter space. In \cite{Bambhaniya:2016rbb} the authors focused particularly on resonant leptogenesis \cite{Pilaftsis:2003gt} which allows the mass scale of the heavy Majorana neutrinos to be lowered thereby reducing the corrections to the Higgs mass\footnote{The heavy Majorana neutrinos induce quantum corrections to the Higgs mass and without any new states, \emph{e.g.} supersymmetric partners, such corrections can be large.}. It has been shown that for heavy neutrino masses $M_N\lesssim 10^7$~GeV, such radiative corrections are small ($\delta m_h^2 < m_h^2$) \cite{Vissani:1997ys,Farina:2013mla,Bambhaniya:2016rbb}. However, we choose to remain agnostic about the level of fine tuning in Nature in view that there has not been any experimental evidence for new physics around the electroweak scale, generally predicted by extensions of the SM which intend to address Higgs naturalness, and hence do not consider this issue any further. There also have been recent similar analyses completed in the context of  the inverse seesaw mechanism, \emph{e.g.} in Refs.~\cite{Khan:2012zw, Rose:2015fua, Lindner:2015qva, DiLuzio:2017tfn}.\\
  
The paper is organised as follows: in \secref{sec:thermallepto} we present an overview of high-scale thermal leptogenesis and numerically solve the density matrix equations to compute the baryon asymmetry. We describe the procedure for our numerical scan in \secref{sec:numerical} and present the RG analysis and our results in \secref{sec:Results}. Finally, in \secref{sec:Conclusions} we present our conclusions.\\

\section{High-scale Thermal Leptogenesis}
\label{sec:thermallepto}

The mechanism of thermal leptogenesis \cite{Fukugita:1986hr} is partly motivated by the observation of small, but non-zero neutrino masses and its possible connection to the seesaw mechanism. In its simplest formulation, the type-I seesaw mechanism \cite{GellMann:1980vs,Yanagida:1979as,Minkowski:1977sc} introduces singlet Majorana fermions to the SM particle spectrum. From the seesaw relation, the scale of the heavy Majorana neutrinos suppresses the light neutrino masses. Moreover, Sakharov's three conditions are inherently  satisfied within this framework: \textbf{(i)} the heavy Majorana neutrino mass matrix violates lepton number, \textbf{(ii)} the Yukawa matrix which couples the heavy Majorana neutrino to the SM particles can be complex and provide new sources of CP-violation and \textbf{(iii)} finally, due to the expansion of the Universe, the heavy Majorana neutrino is a good candidate to decay out of thermal equilibrium. \\

The lepton asymmetry is  dynamically produced from the CP-violating decays of the heavy Majorana neutrinos. There are also washout processes present which compete with these decays and act to reduce the 
overall asymmetry. The final lepton asymmetry is  partially reprocessed to a baryon asymmetry via weak sphaleron processes which proceed at unsuppressed rates above the electroweak scale \cite{Khlebnikov:1988sr}. 
The leptogenesis era  occurs approximately when the temperature of the Universe equals the mass scale of the decaying heavy Majorana neutrino  ($T \approx M_1$). \\

In general, it was thought thermal leptogenesis, in the context of a type-I seesaw, was not viable below 
 $M_1\approx10^{9}$ GeV \cite{Davidson:2002qv}. This lower limit, otherwise known as the Davidson-Ibarra bound, is derived from the requirement that the CP-asymmetry
 parameter $\epsilon$ is sufficiently large to produce the observed baryon asymmetry assuming the efficiency factor (which is used to parametrise the washout) is approximately unity. 
However, this bound may be lowered to $M_{1}\approx 10^{6}$ GeV if one foregoes $\gtrsim\mathcal{O}\left(10 \right)$ fine-tuning to the light neutrino masses \cite{Moffat:2018wke}.
Using  $\Delta L = 2$ washout processes, due to scattering mediated by off-shell $N_2$ or $N_3$, an upper bound $M_{1}\lesssim 10^{14}$ GeV \cite{Agashe:2018oyk} was found. However, a more refined calculation, which 
includes $\Delta L = 1$ processes,  further extends this upper bound to $M_{1}\lesssim 10^{15}$ GeV \cite{Buchmuller:2004nz,Buchmuller:2002rq}. Therefore, there is a large energy-scale window for non-resonant thermal leptogenesis ranging from $10^{6}~{\rm GeV}\lesssim M_1\lesssim10^{15}$~GeV.\\

To implement a type-I seesaw mechanism the following terms are added to the SM Lagrangian
\begin{equation}\label{eq:lag}
\mathcal{L} = i\overline{N_{i}}\slashed{\partial}N_{i}  -{Y_N}_{\alpha i}\overline{L_{\alpha}}\tilde{H}N_{i}-\frac{1}{2}M_{i}\overline{N^c_{i}}N_{i} + \text{h.c.},
\end{equation}
where $\alpha=e, \mu, \tau$, $i= 1, 2, 3$, $Y_N$ is the Yukawa matrix, $H$ the Higgs $\SU (2)_L$ doublet, $\tilde{H} = i \sigma_2 H^*$ and $L^T = \left( \nu^T_L, l^T_L  \right)$ is the leptonic $\SU (2)_L$ doublet.
The connection between thermal leptogenesis, or in fact any mechanism of leptogenesis which implements a type-I seesaw, and neutrino observables can be expressed by the Casas-Ibarra parametrisation \cite{Casas:2001sr} for the matrix $Y_N$
\begin{equation}\label{matrix}
Y_N=\frac{1}{v}U\sqrt{m_\nu}\,R^T\sqrt{M_N},
\end{equation}
where $v=174$~GeV is the vacuum expectation value of the Higgs, $m_\nu$ ($M_N$) the diagonal mass matrices of light (heavy) neutrinos and $R$ is an orthogonal, complex matrix.
 The  Pontecorvo-Maki-Nakagawa-Sakata (PMNS) 
 matrix, $U$,  describes the misalignment between the mass and flavour basis of the active neutrinos and 
 contains six real parameters: three measured mixing angles and three CP-phases. The Dirac phase, $\delta$,
 enters the neutrino oscillation probabilities subdominantly and remains largely unconstrained by experimental data.
 The Majorana phases  \cite{Bilenky:1980cx,Schechter:1980gr}, which are physical if and only if neutrinos are Majorana particles, 
 ($\alpha_{21}, \alpha_{31} = [0, 4\pi]$\footnote{This range is necessary as the Majorana phases  enter into the expression for the neutrino mixing matrix in the form $e^{\frac{i\alpha_{21}}{2}}$ and $e^{\frac{i\alpha_{31}}{2}}$ \cite{Molinaro:2008rg}.})  cannot be determined using neutrino oscillation experiments and have to be measured using alternative methods. \\
 
In this work, we take the experimental central values for the active neutrino masses, $m_{\nu_i}$, the mixing angles, $\theta_{ij}$, and the Dirac CP-phase, $\delta$, which are given by \cite{Esteban:2016qun} 
\begin{equation}\label{eq:numvalues}
\begin{aligned}
\theta_{13} = 8.52^\circ, \hspace{5mm} \theta_{12} = 33.63^\circ, \hspace{5mm} 
\theta_{23}=48.7^\circ, \hspace{5mm} 
\delta = 228 ^\circ, \notag \\
\Delta m^2_{21}  = 7.40\times 10^{-5} \, \text{eV}^2, \hspace{5mm} 
\Delta m^2_{31} = 2.515 \times 10^{-3} \, \text{eV}^2 , 
\end{aligned}
\end{equation}
for a normally ordered mass spectrum. 
 Moreover, as the two Majorana phases are currently completely unconstrained, we fix their values to be CP-conserving, $\alpha_{21}=\alpha_{31}=0$. 
In addition to the low-energy CP-phases, there are also phases associated with the  $R$ matrix. As the $R$ matrix is complex and orthogonal, it may be parametrised in terms of three complex rotation matrices
\begin{equation}\label{eq:Rmatr}
R=\begin{pmatrix}
1 & 0 & 0 \\
0 & c_{\theta_{1}} & s_{\theta_{1}} \\
0 &- s_{\theta_{1}} & c_{\theta_{1}} 
\end{pmatrix}
\begin{pmatrix}
c_{\theta_{2}} & 0 & s_{\theta_{2}} \\
0 & 1 & 0\\
-s_{\theta_{2}} & 0 & c_{\theta_{2}} 
\end{pmatrix}
\begin{pmatrix}
c_{\theta_{3}} & s_{\theta_{3}} & 0\\
-s_{\theta_{3}} & c_{\theta_{3}} & 0\\
0 & 0 & 1
\end{pmatrix},
\end{equation}
where $c_{\theta_{i}}=\cos\theta_{i}$, $s_{\theta_{i}}=\sin\theta_{i}$ for $i=1,2,3$; and $\theta_{i}$ are complex angles. 
Therefore, in general, the baryon asymmetry produced from thermal leptogenesis is a function of 18 parameters of the Casas-Ibarra parametrisation. \\
\subsection{Leptogenesis with Two Heavy Majorana Neutrinos}

In order to simplify our discussion, we  consider the decoupling limit of the heaviest Majorana neutrino, $N_{3}$.  
From the seesaw relation, this causes the lightest neutrino to be massless, which is consistent with data from oscillation experiments. In this limit, the $R$ matrix assumes the following simplified form \cite{Ibarra:2003up,Antusch:2011nz,Blanchet:2007qs}
\begin{equation}
R^{\rm NO} = 
\begin{pmatrix}
0 & \cos \theta& \pm\sin \theta \\
0 & \pm\sin \theta & \cos\theta \\
1& 0 & 0
\end{pmatrix},
\quad 
R^{\rm IO} = \begin{pmatrix}
 \cos \theta & \pm\sin \theta & 0\\
\pm\sin \theta& \cos \theta & 0\\
0& 0 & 1
\end{pmatrix},
\end{equation}
for normal (NO) and inverted (IO) ordering respectively. As a consequence, the Yukawa matrix, for normal ordering, may be written as
\begin{equation}\label{eq:Yukawa}
\begin{aligned}
Y_N=&\frac{1}{v}  \begin{pmatrix} U_{e1} & U_{e1} &U_{e3}  \\ 
 U_{\mu1} & U_{\mu2} &U_{\mu3}  \\ 
 U_{\tau1} & U_{\tau2} &U_{\tau3}  
\end{pmatrix} 
\begin{pmatrix}0 & 0&0\\ 
0 & \sqrt{m_2} &0  \\ 
0 & 0&\sqrt{m_3}
\end{pmatrix} 
\begin{pmatrix} 0 & c_{\theta} &s_{\theta}  \\ 
0 & -s_{\theta} &c_{\theta} \\ 
1 & 0 &0 
\end{pmatrix} ^{T}
\begin{pmatrix} \sqrt{M_1} & 0&0\\ 
0 & \sqrt{M_2} &0  \\ 
0 & 0&0
\end{pmatrix} \\
=&\frac{1}{v}
  \begin{pmatrix}
 \sqrt{M_{1}} \left(U_{\text{e2}}\sqrt{m_{2}} c_{\theta } +U_{\text{e3}}\sqrt{m_{3}} s_{\theta } \right)\;&\;\sqrt{M_{2}}
   \left(U_{\text{e3}}\sqrt{m_{3}} c_{\theta } -U_{\text{e2}}\sqrt{m_{2}} s_{\theta } \right) \;&\; 0 \\
 \sqrt{M_{1}} \left(U_{\text{$\mu $2}}\sqrt{m_{2}} c_{\theta } +U_{\text{$\mu $3}}\sqrt{m_{3}} s_{\theta } \right) \;&\; \sqrt{M_{2}}
   \left(U_{\text{$\mu $3}}\sqrt{m_{3}} c_{\theta } -U_{\text{$\mu $2}}\sqrt{m_{2}} s_{\theta } \right)\;&\; 0 \\
\sqrt{M_{1}} \left(U_{\text{$\tau $2}}\sqrt{m_{2}} c_{\theta } +U_{\text{$\tau $3}}\sqrt{m_{3}} s_{\theta } \right) \;&\;\sqrt{M_{2}}
   \left(U_{\text{$\tau $3}}\sqrt{m_{3}} c_{\theta } -U_{\text{$\tau $2}}\sqrt{m_{2}} s_{\theta } \right) \;&\; 0 \\
\end{pmatrix},
\end{aligned}
\end{equation}
where $\theta$ is a complex angle. 
The scenario of thermal leptogenesis with two heavy Majorana neutrinos is particularly appealing as the dimensionality of the
model parameter space is greatly reduced (from 18 to 11) and for this reason has been investigated using the unflavoured approximation \cite{Frampton:2002qc,King:2002qh,Chankowski:2003rr,Ibarra:2003up} and also including flavour effects \cite{Abada:2006ea,Antusch:2006cw,Molinaro:2008cw,Anisimov:2007mw}. 
As we have assumed the best-fit values for the low-energy neutrino parameters, the 
 lepton asymmetry depends only on four parameters ($M_{1}$, $M_{2}$, ${\rm Re}[\theta]$ and ${\rm Im}[\theta]$) which we assume take the ranges shown in \tabref{tab:ranges}.
It is worth noting that as we assume a strictly hierarchical mass spectrum for the heavy neutrinos ($M_{2}\geq 10M_{1}$)  we find  the lower range for viable 
leptogenesis to be $M_{1}\approx 10^{9}$ GeV. Although non-resonant thermal leptogenesis is possible at lower scales, a milder mass hierarchy is required.\\

\begingroup
\setlength{\tabcolsep}{10pt} 
\renewcommand{\arraystretch}{1.5} 
\begin{table}[t]
\centering
\begin{tabular}{ | l | l | }
\hline
 Parameter  &  Scan Range  \\
 \hline 
$M_1$ & $[10^{9}$, $10^{15}]$ GeV \\
$M_2$ & $[10^1$, $10^{4}] \, M_{1}$ \\
Re$[\theta]$	     & $\left[ -\pi, \pi  \right]$ \\
Im$[\theta]$          & $\left[ -\pi, \pi  \right]$ \\
\hline
\end{tabular}\caption{Ranges of the four free parameters in our numerical scan.}\label{tab:ranges}
\end{table}
\endgroup

\subsection{Density Matrix Equations}
In order to calculate the time evolution of the  lepton asymmetry we solve the density matrix equations detailed in  \cite{Blanchet:2011xq} where the theoretical background and derivation of these kinetic equations are presented. Due to our assumption of a hierarchical heavy neutrino mass spectrum, the contribution to the lepton asymmetry from $N_{2}$ is negligible and the density matrix equation are given by
\begin{align}\label{eq:full3}
\frac{dn_{N_{1}}}{dz}=&-D_{1}(n_{N_{1}}-n^\text{eq}_{N_{1}}),\\
 \frac{dn_{\alpha\beta}}{dz} =&~\epsilon^{(1)}_{\alpha\beta}D_{1}(n_{N_{1}}-n^\text{eq}_{N_{1}})-\frac{1}{2}W_{1}\left\{P^{0(1)},n\right\}_{\alpha\beta}  \nonumber\\
-&\frac{{\rm Im}(\Lambda_{\tau})}{Hz}\left[\begin{pmatrix}1&0&0\\ 0&0&0 \\
0&0&0 \end{pmatrix},\left[\begin{pmatrix}1&0&0\\ 0&0&0 \\
0&0&0 \end{pmatrix},n\right]\right]_{\alpha\beta}
-\frac{{\rm Im}(\Lambda_{\mu})}{Hz}\left[\begin{pmatrix}0&0&0\\ 0&1&0 \\
0&0&0 \end{pmatrix},\left[\begin{pmatrix}0&0&0\\ 0&1&0 \\
0&0&0 \end{pmatrix},n\right]\right]_{\alpha\beta} \nonumber,
\end{align}
where  Greek letters denote flavour indices, $n_{N_{1}}$ ($n^\text{eq}_{N_{1}}$) is the abundance (equilibrium abundance) of the lightest heavy Majorana neutrino \footnote{This quantity is normalised to a co-moving volume containing one right-handed neutrino which is ultra-relativistic and in thermal equilibrium.},  $D_1$ ($W_1$) denotes the decay (washout) of $N_{1}$. Moreover, $H$ denotes the Hubble expansion rate and $\Lambda_{\alpha}$ is the self-energy of $\alpha$-flavoured leptons. The thermal widths, $\text{Im}\left(\Lambda_{\alpha}\right)$, of the charged leptons is given by the imaginary part of the self-energy correction to the lepton propagator in the thermal plasma. 
The projection matrices, $P^{0(i)}_{\alpha \beta}$  describe how a given flavour of lepton is washed out and the CP-asymmetry matrix details the decay asymmetry generated by  $N_{1}$ and is denoted by  $\epsilon^{(1)}_{\alpha\beta}$.  
The density matrix equations of    \equaref{eq:full3} may be used to calculate the lepton asymmetry in all flavour regimes and accurately describes the transitions between them \cite{Barbieri:1999ma, Abada:2006fw,DeSimone:2006nrs,Blanchet:2006ch,Blanchet:2011xq}.  The complex off-diagonal entries of $n_{\alpha \beta}$ allow for a quantitative description of these transitions. For example, if leptogenesis occurs at temperatures $10^{9}~{\rm GeV}\lesssim T \lesssim 10^{12}~{\rm GeV}$,  the term ${\rm Im}\left(\Lambda_{\tau}\right)/Hz$ damps the evolution of the off-diagonal elements of $n_{\alpha \beta}$. This  reflects the loss of coherence of the tau charged lepton state as the SM tau Yukawa couplings come into thermal equilibrium.  Additional formulae and further discussion of the density matrix equations in \equaref{eq:full3} may be found in \appref{sec:DMEapp}.

\section{Numerical Procedure}
\label{sec:numerical}
Our goal in this work is to ascertain  whether  or not
a minimal extension of the SM, which simultaneously explains neutrino masses and the observed BAU, can be valid up to the Planck scale without the requirements of additional new degrees of freedom.
To investigate this, we first fix the measured central values in the active neutrino sector as given in \equaref{eq:numvalues} and then perform a random scan of the four-dimensional leptogenesis model parameter space as shown in \tabref{tab:ranges}\footnote{Note that this procedure is undertaken for normally ordered active neutrinos. The results are qualitatively similar for inverted ordering.}. For a given point in this space we calculate a Yukawa matrix, $Y_N$, as parametrised in \equaref{eq:Yukawa}. Subsequently, we numerically solve the density matrix equations of \equaref{eq:full3} to find the baryon-to-photon ratio associated to a particular point in the model parameter space. The  baryon-to-photon ratio is defined as
\begin{equation}
\eta_B \equiv \frac{n_B-n_{\overline{B}}}{n_\gamma},
\end{equation}
where $n_B$, $n_{\overline{B}}$ and $n_\gamma$ are the number densities of baryons, antibaryons and photons respectively.
This quantity has been measured from Big-Bang nucleosynthesis (BBN) \cite{Cooke:2013cba,Patrignani:2016xqp} and 
Cosmic Microwave Background (CMB) radiation data  \cite{Ade:2015xua}:
\[
\begin{aligned}
{\eta_{B}}_{\text{BBN}}  & = \left(5.80-6.60\right)\times 10^{-10}, \\
{\eta_{B}} _{\text{CMB}}& = \left(6.02-6.18\right)\times 10^{-10},
\end{aligned}
\]
at 95$\%$ CL, respectively. For a given point in the model parameter space, we define \emph{successful leptogenesis}
to produce $\eta_B>5.8\times10^{-10}$.\\

Once we have a subset of points that lead to successful leptogenesis, we subsequently solve the RG equations for all the couplings in the model and determine whether  they remain perturbative and the electroweak vacuum  stable or metastable up to the Planck scale\footnote{Similar works that study how the electroweak vacuum is affected by new physics related to the origin of neutrino masses include \cite{Gogoladze:2008gf,He:2012ub,Chakrabortty:2012np,Kobakhidze:2013pya,Bonilla:2015kna,Ng:2015eia,Haba:2016zbu,Garg:2017iva}.}, $M_{\rm PL}\!=\!1.22\times 10^{19}$ GeV.  We detail this further in \secref{sec:Results}.


\section{Results}
\label{sec:Results}
In this Section, we present the results of our RG analysis of high-scale leptogenesis for normally ordered light neutrino mass spectrum. We consider the Lagrangian in \equaref{eq:lag}, which augments the SM particle spectrum by three heavy Majorana neutrinos that are responsible for the masses of active neutrinos and the baryon asymmetry of the Universe.
We determine the UV-robustness of the viable parameter space in this minimal case, as discussed, and the non-standard scenario of asymptotically safe QCD.\\

\subsection{Vacuum Stability and Perturbativity in High-Scale Leptogenesis}
\label{sec:SM+N}

There are two problems that can arise when extrapolating the coupling constants of the model to high scales. \textbf{(i)} Dimensionless coupling constants in the Lagrangian could grow with energy such that they become larger than $\mathcal{O}(1)$, rendering the theory non-perturbative. \textbf{(ii)} Adding new fermionic degrees of freedom gives a negative contribution to the $\beta$-function of the Higgs quartic and this can drive the already-metastable electroweak vacuum to violate the metastability bound. \\

In order to study the stability of the electroweak vacuum we use the one-loop RG-improved effective potential for the Higgs. For the RG analysis we solve the two-loop $\beta$-functions for the SM couplings that can be found in \cite{Buttazzo:2013uya} and one-loop for $Y_N$, both computed in the ${\rm \overline{MS}}$ scheme. For simplicity we provide the one-loop RG equations \cite{Cheng:1973nv} below. 
\begin{itemize}
\item Gauge couplings \cite{Gross:1973id,Politzer:1973fx} :
\end{itemize}
\begin{align} \label{eq:betagauge} 
(4\pi)^2 \frac{dg_Y}{d\ln \mu} & = \frac{41}{6}g_Y^3, \hspace{10mm}
(4\pi)^2 \frac{dg_2}{d\ln \mu} = -\frac{19}{6}g_2^3, \hspace{10mm}
(4\pi)^2 \frac{dg_3}{d\ln \mu}  = -7 g_3^3.
\end{align}

\begin{itemize}
\item The Higgs quartic, the top Yukawa and the neutrino Yukawa matrix \cite{Grzadkowski:1987tf, Pirogov:1998tj,Casas:1999cd,Antusch:2002rr}:
\end{itemize}
\begin{align}
(4\pi)^2 \frac{d\lambda_H}{d\ln \mu} & = \lambda_H \left( 24\lambda_H + 12 y_t^2 - 9 g_2^2 - 3 g_Y^2 \right)
                                       - 6y_t^4 + \frac{9}{8}g_2^4 + \frac{3}{8} g_Y^4 + \frac{3}{4} g_2^2 g_Y^2 \,  \nonumber \\
                                       & \phantom{=} + 4  {\rm Tr}[Y_N^{\dagger} Y_N] \lambda_H   - 2 {\rm Tr}[ (Y_N^{\dagger} Y_N )^2  ], \label{eq:betaLH} \\[2ex]
(4\pi)^2 \frac{dy_t}{d\ln \mu}     & = y_t \left( \frac{9}{2}y_t^2 - \frac{17}{12} g_Y^2 - \frac{9}{4} g_2^2 - 8 g_3^2   \right) + y_t {\rm Tr}[ Y_N^{\dagger} Y_N ] , \label{eq:betayt} \\[2ex]
(4\pi)^2 \frac{d Y_N}{ d \ln \mu} & =  \frac{3}{2} Y_N Y_N^{\dagger} Y_N +  Y_N {\rm Tr}[Y_N^{\dagger} Y_N] - Y_N \left( \frac{3}{4}{g_Y^2} + \frac{9}{4} g_2^2 - 3 y_t^2 \right), \label{eq:betaYN}
\end{align}
where $\mu$ corresponds to the RG scale and the Higgs quartic $\lambda_H$ is defined by writing the SM Higgs potential as $V=-\frac{1}{2}M_h^2 |H|^2 + \lambda_H |H|^4$. The contribution from each column $i$ of the Yukawa matrix to the $\beta$-functions is set to zero when $\mu<M_i$. From the last term in Eq.~\eqref{eq:betaLH}, we observe that $Y_N$ gives a negative contribution to the running of $\lambda_H$. This implies that large Yukawa couplings will destabilise the electroweak vacuum. Furthermore the positive contribution from $Y_N$ to $\beta_{y_t}$ in Eq.~\eqref{eq:betayt} can drive the top Yukawa to large positive values which further pushes $\lambda_H$ below the metastability bound.
We do not consider RG running effects on the parameters of the active neutrino sector as in general such effects are weak \cite{Casas:1999tg,Antusch:2003kp}.\\


We impose the following perturbativity constraints on the SM and the neutrino Yukawa couplings
\begin{equation}\label{eq:perturbativity}
g_i(\mu), \, y_t(\mu), \, \lambda_H(\mu) \leq \sqrt{4\pi}, \hspace{10mm}|Y_N^{ij}(\mu)|\leq \sqrt{4\pi},  \hspace{10mm} {\rm Tr}[Y_N^\dagger(\mu) Y_N(\mu)] \leq 4\pi.
\end{equation}
The effective Higgs quartic coupling can be defined in terms of the one-loop contributions to the Higgs potential and is given as follows \cite{Casas:1994qy, Buttazzo:2013uya},
\begin{align}\label{eq:lambdaeff}
\lambda_{\rm eff}& = \lambda_H+  \frac{1}{16\pi^2} \left[ \frac{3}{16} \left(g_Y^2 + g_2^2\right)^2 \left( \ln \frac{g_Y^2 + g_2^2}{4} - \frac{5}{6} \right)  + \frac{3}{8} g_2^4 \left( \ln \frac{g_2^2}{4} - \frac{5}{6} \right) \right. \nonumber \\[1ex]
& \phantom{=} \left.  - \, 3 y_t^4 \left( \ln \frac{y_t^2}{2} - \frac{3}{2} \right) + 3 \lambda_H^4 \left( 4 \ln \lambda_H - 6 + 3\ln 3 \right) \right] + \lambda_N,
\end{align}
where the $\mu$ dependence of each coupling is not explicitly written and we have taken $\mu$ equal to the value of the Higgs field in order to minimize the contribution from the logarithms. The last term in the expression above corresponds to the contribution from the heavy Majorana neutrinos that only enters at $\mu\geq M_{N_i}$. We consider a leptogenesis scenario where the heaviest sterile neutrino $N_3$ is decoupled; for the RG analysis we set $M_3$ close to the Planck scale. The contribution from the neutrino Yukawa couplings to the effective Higgs quartic coupling at one-loop is given as
\begin{equation}
\lambda_N = -  \frac{1}{32\pi^2} \left[(Y_N^\dagger Y_N)^2_{ii} \left\{ \ln \frac{(Y_N^\dagger Y_N)_{ii}}{2} - 1 \right\} +  (Y_N^\dagger Y_N)^2_{ii} \left\{ \ln \frac{(Y_N^\dagger Y_N)_{ii}}{2} - 1 \right\}  \right],
\end{equation}
where $Y_N^\dagger Y_N$ is written in the diagonal basis \cite{Khan:2012zw}.\\

\begin{figure}[t]
\centering
\includegraphics[width=.485\textwidth]{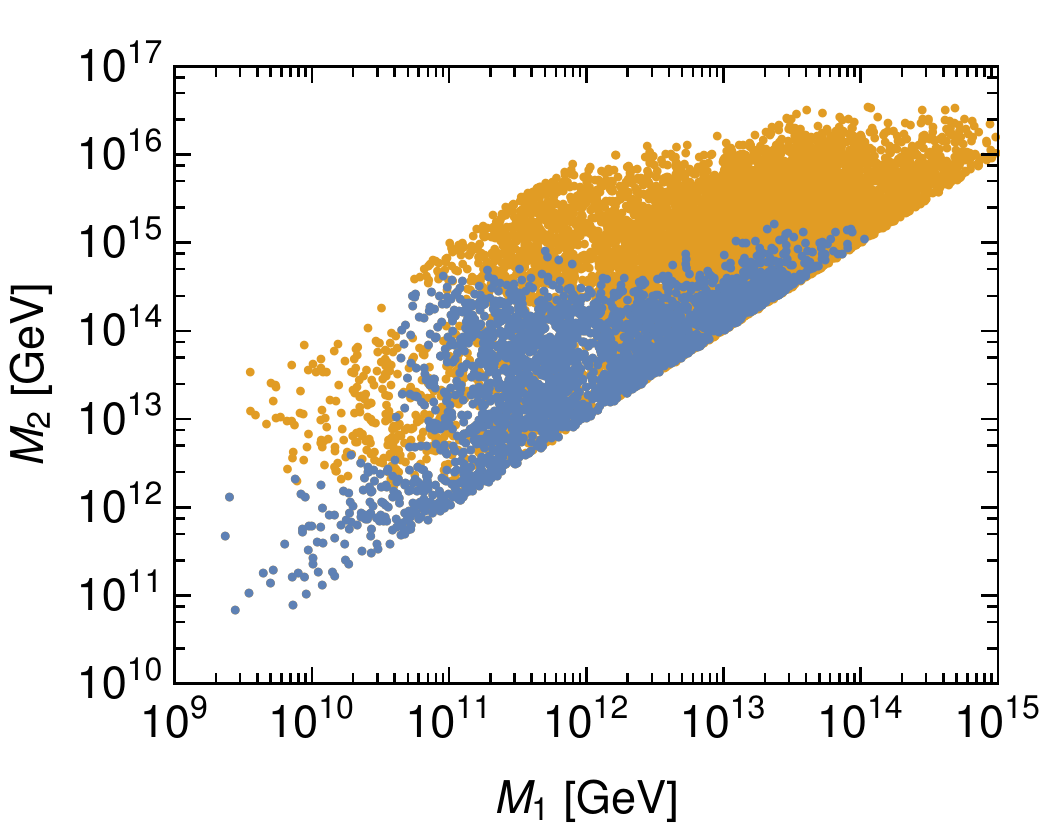} \,\,
\includegraphics[width=.485\textwidth]{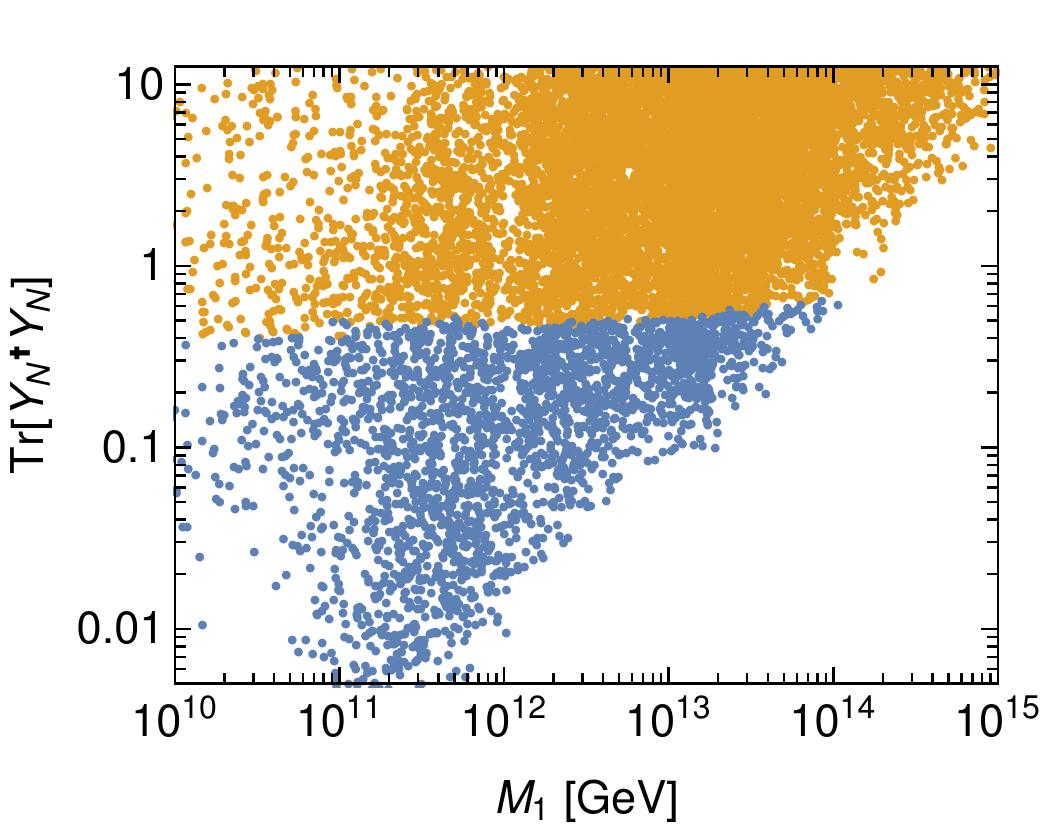}
\caption{Scatter plot of all the points in our scan that lead to successful leptogenesis. Points in orange correspond to the points where any of the Lagrangian coupling constants become non-perturbative below the Planck scale and/or lead to a metastable EW vacuum with a lifetime smaller than the age of the Universe. Points in blue satisfy these constraints up to the Planck scale as well as leading to successful leptogenesis.}
\label{fig:scatter1}
\end{figure}

A detailed RG study of the Higgs potential in the SM has shown that at high scales the quartic coupling turns negative and the potential develops a new minimum \cite{Degrassi:2012ry,Buttazzo:2013uya}, for earlier related work on this topic see \cite{Lindner:1985uk, Sher:1988mj}. Taking the central values for measurements of the Higgs and the top mass, this secondary vacuum is deeper than the electroweak vacuum and hence, given sufficient time, there is a chance the Higgs field will tunnel into it. Requiring the age of the Universe to be larger than the lifetime of the electroweak vacuum implies a lower bound for negative values of $\lambda_{\rm eff}(\mu)$ \cite{Buttazzo:2013uya}
\begin{equation}\label{eq:lambdastab}
|\lambda_{\rm eff}(\mu)| > \frac{2\pi^2}{3} \frac{1}{\ln(\tau \mu)},
\end{equation}
where $\tau=4.35\times 10^{17}$ s is the age of the Universe \cite{Ade:2015xua}. Using the one-loop RG improved Higgs potential and the two-loop $\beta$-functions with the most recent measurements for the top and the Higgs mass we find that $\lambda_{\rm eff}$ becomes negative at $\mu_{\rm inst}\!\approx \! 5.2 \times 10^{11}$ GeV in the SM.\\ 

Although non-perturbatively it is possible to show the gauge-independence of tunneling rates by applying Nielsen identities \cite{Nielsen:1975fs} to the false-vacuum effective action \cite{Plascencia:2015pga}, it appeared that in perturbative calculations the lifetime of the electroweak vacuum displayed gauge-dependence \cite{Patel:2011th, DiLuzio:2014bua, Lalak:2016zlv}. There has been recent progress in performing a perturbative calculation of the decay rate that have resolved these issues \cite{Andreassen:2017rzq, Chigusa:2017dux}. In this work, we use the Landau gauge; however, we anticipate our conclusions not to change for a different choice of gauge-fixing.\\
 
\begin{figure}[t]
\centering
\includegraphics[width=.485\textwidth]{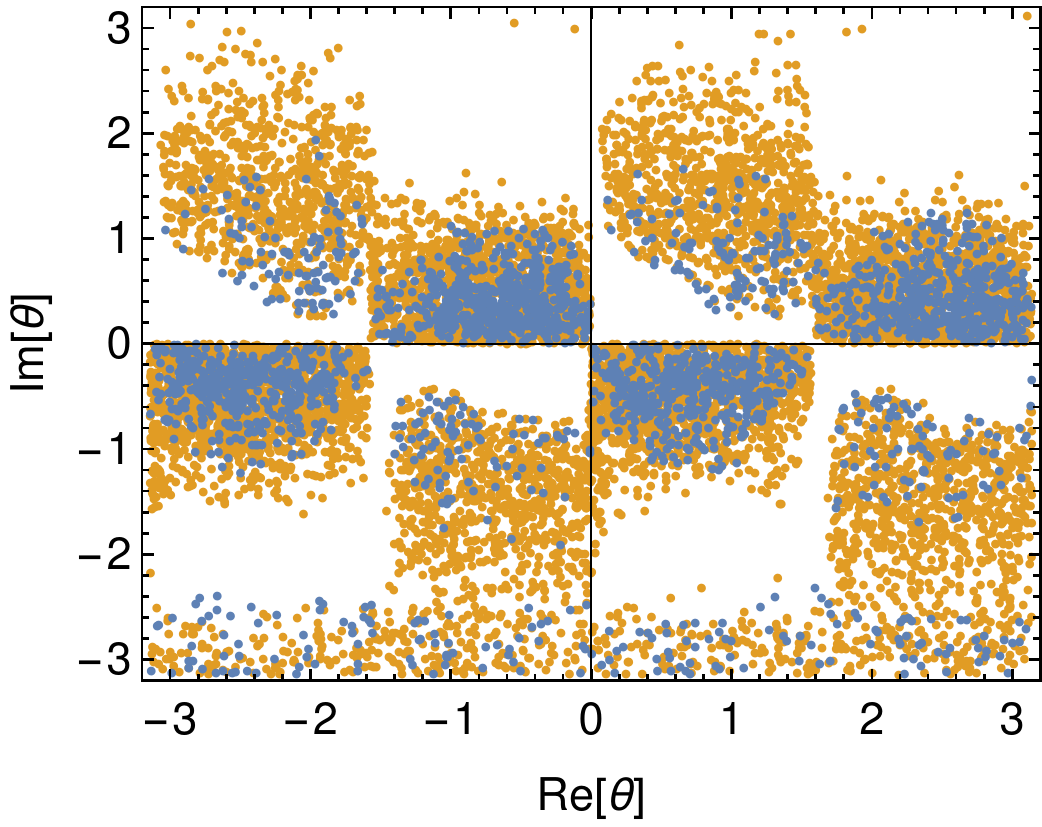} \,\,
\includegraphics[width=.485\textwidth]{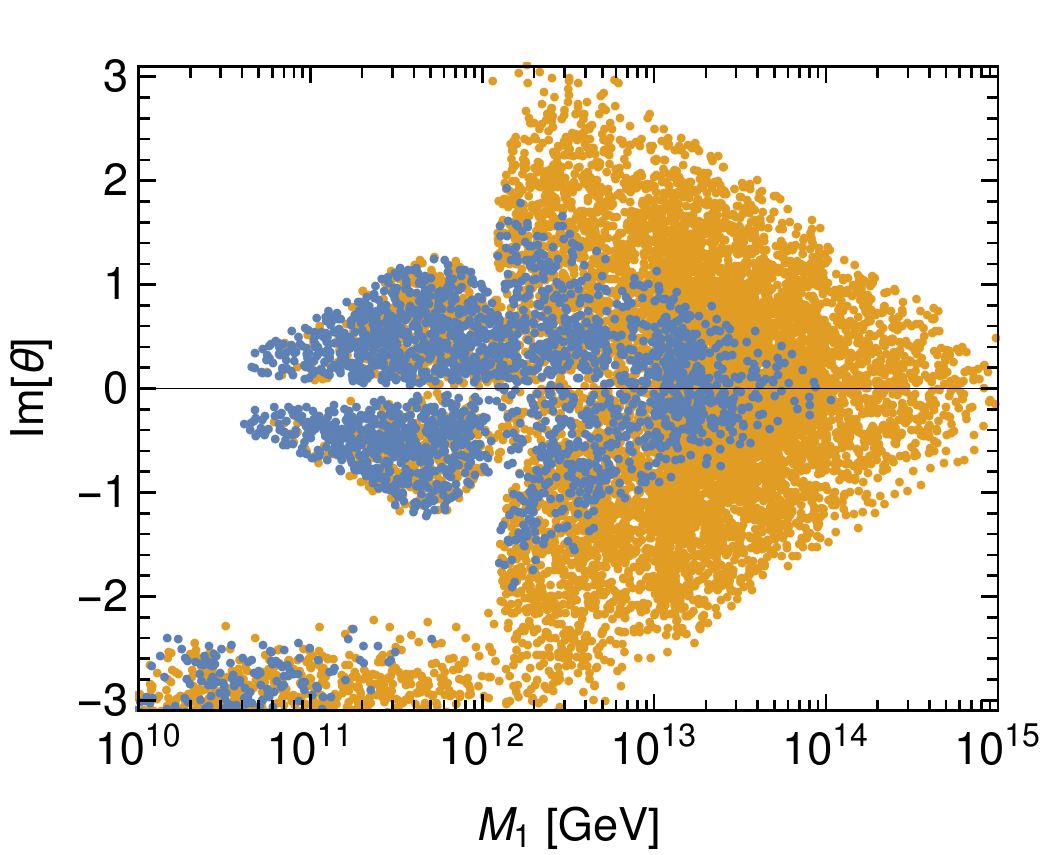}
\caption{The colouring of the points is the same as in Fig.~\ref{fig:scatter1}. We display the dependence on the real and imaginary part of the complex angle that enters in the $R$ matrix.}
\label{fig:scatter2}
\end{figure}

The initial conditions of the RG running are taken at $\mu\!=\!M_t$ and we use the central value of the top quark mass from recent measurements performed by ATLAS and CMS, $M_t\!=\!172.5$ GeV \cite{ATLAS:2017lqh,Menke:11}. We take $M_W\!=\!80.384$ GeV and $M_h\!=\!125.09$ GeV for the pole mass of the $W$ and the Higgs bosons respectively and $\alpha_3(M_Z)\!=\!0.1184$ for the strong coupling constant \cite{Patrignani:2016xqp}. We complete the calculations using  the next-to-next-to-leading order initial values for the SM couplings as given in Ref. \cite{Buttazzo:2013uya} 
\begin{align*}
g_3(\mu\!=\!M_t)          & = 1.1666 + 0.00314\frac{\alpha_3(M_Z)-0.1184}{0.0007} -0.00046 \left( \frac{M_t}{\text{GeV}} - 173.1 \right),  \\[2ex]
g_2(\mu\!=\!M_t)          & = 0.64779 + 0.00004 \left( \frac{M_t}{\text{GeV}} - 173.34 \right) + 0.00011 \frac{ M_W - 80.384 \text{ GeV}}{0.014 \text{ GeV}},  \\[2ex]
g_Y(\mu\!=\!M_t)           & = 0.35830 + 0.00011 \left( \frac{M_t}{\text{GeV}} - 173.34 \right) - 0.00020\frac{ M_W - 80.384 \text{ GeV}}{0.014 \text{ GeV}}, 
\end{align*}
\begin{align*}
y_t(\mu\!=\!M_t)          & = 0.93690 + 0.00556 \left( \frac{M_t}{\text{GeV}} - 173.34 \right) -0.00042 \frac{\alpha_3(M_Z)-0.1184}{0.0007},\\[2ex]
\lambda_H (\mu\!=\!M_t)   & = 0.12604 + 0.00206 \left( \frac{M_h}{\text{GeV}} - 125.15 \right)  -0.00004  \left( \frac{M_t}{\text{GeV}} - 173.34 \right).
\end{align*}

\begin{figure}[t]
\centering
\includegraphics[width=.48\textwidth]{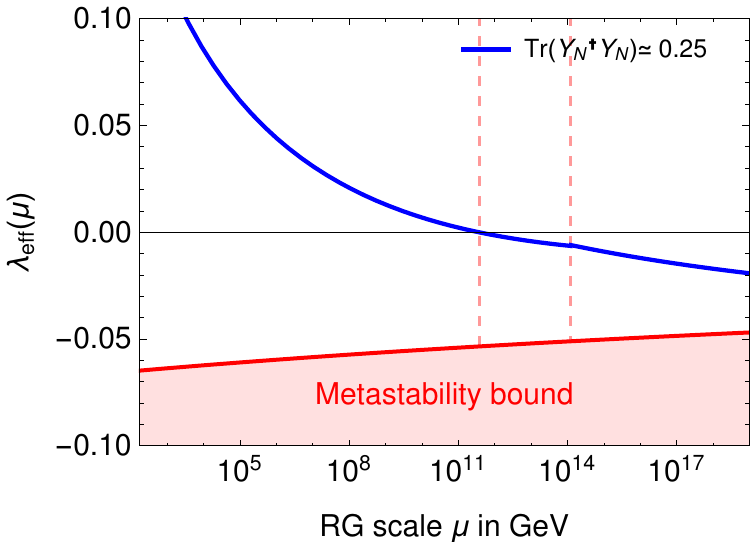} \, \,
\includegraphics[width=.48\textwidth]{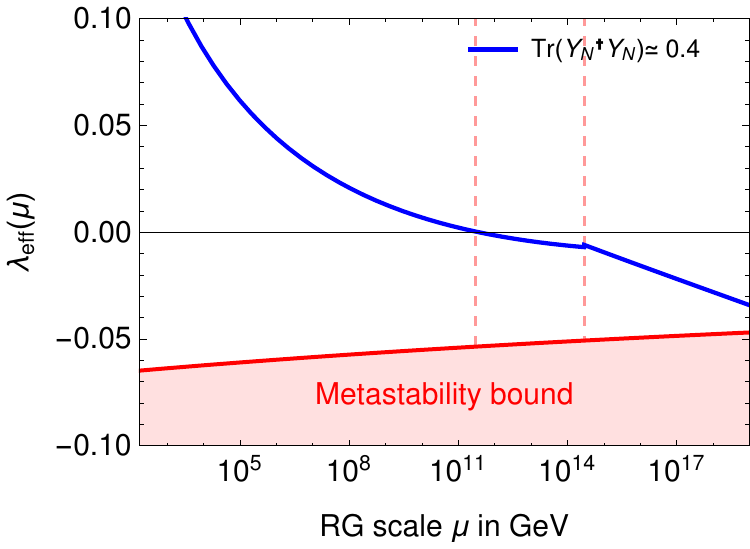}
\includegraphics[width=.48\textwidth]{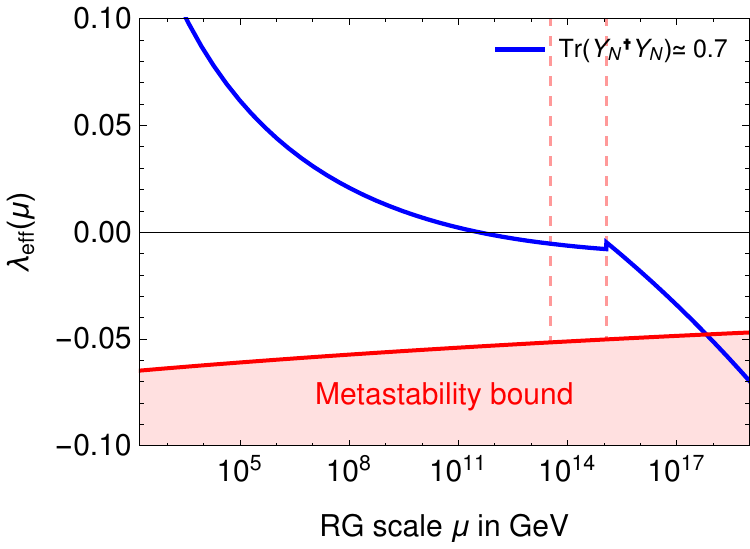} \, \,
\includegraphics[width=.48\textwidth]{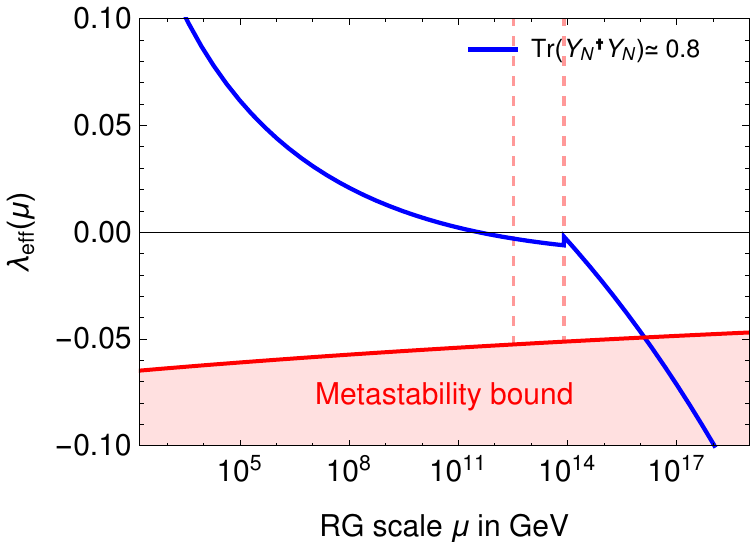}
\caption{RG running of the effective Higgs quartic coupling as a function of the RG scale for points that lead to successful leptogenesis in the type-I seesaw mechanism. The red shaded region corresponds to the region where the lifetime of the EW vacuum is smaller than the age of the Universe. The dashed vertical lines correspond to the masses of $N_1$ and $N_2$. We show the running for different values of ${\rm Tr}[Y_N^\dagger Y_N]$ at the seesaw scale.}
\label{fig:lambdafig}
\end{figure}

We present our results in Figs.~\ref{fig:scatter1}-\ref{fig:scatter2}. As can be seen in the left panel of Fig.~\ref{fig:scatter1}, the region with smaller $M_i$ is less populated compared to larger $M_i$. This is expected since the CP-asymmetry in the decays of the heavy Majorana neutrino is proportional to its mass, $M_{1}$.
As we require the model to be UV-robust up to the  Planck scale we find an upper bound on the lightest right-handed Majorana neutrino $M_1\!\lesssim \!1.2\times10^{14}$ GeV, this is similar to the one found in RG studies of the type-I seesaw without considering leptogenesis \cite{Casas:1999cd,EliasMiro:2011aa}. We also find an upper bound on the  the mass splitting for $M_1>10^{13}$ GeV, namely $\Delta M_{12} / M_1\!\equiv\!(M_2-M_1)/M_1\lesssim 10^2$ must be satisfied. The bounds we provide are conservative, in the sense that by taking a non-zero mass for the lightest active neutrino they will only become stronger. This is due to the seesaw relation in which larger values for $m_1$ require larger Yukawa couplings.\\ 

The right panel in Fig.~\ref{fig:scatter1} shows a scatter plot of $M_{1}$ versus ${\rm Tr}[Y_N^\dagger Y_N]$ where the values for the latter are computed at the seesaw scale. This plot shows that there is an upper bound ${\rm Tr}[Y_N^\dagger Y_N]\lesssim 0.66$. This bound does  not significantly  differ from the one found in Refs.~\cite{Khan:2012zw,Rose:2015fua} for the low-scale inverse seesaw mechanism. This is because neutrino Yukawas $\sim 0.5$ have a drastic effect in driving $\lambda_{\rm eff}$ below the metastability bound. In Fig.~\ref{fig:lambdafig} we present the running of $\lambda_{\rm eff}(\mu)$ for different values of ${\rm Tr}[Y_N^\dagger Y_N]$. As can be seen, for larger values of the neutrino Yukawa couplings, the effective Higgs quartic quickly crosses into the metastability bound. 
\\

In the left panel of Fig.~\ref{fig:scatter2}, we display the points leading to successful leptogenesis in the
$\text{Im}[\theta]$--$\text{Re}[\theta]$ plane. The periodic behaviour in $\text{Re}[\theta]$ may be inferred as the 
analytic forms of the 
washout, projection matrices and CP-asymmetries are all trigonometric functions of $\text{Re}[\theta]$. This periodic behaviour we observe concurs with the results of  \cite{Antusch:2011nz}. In spite of this periodicity in $\text{Re}[\theta]$, the 
constraints from vacuum stability and perturbativity are not sensitive to the absolute magnitude of this parameter.\\

In the right panel of  Fig.~\ref{fig:scatter2}, we observe that for $M_1 \gtrsim 10^{12}$ GeV, there is a preference for small ${\rm Im}[\theta]$. This is because as $M_{1}$ is increased, ${\rm Im}[\theta]$ cannot grow too large 
(given that Yukawa couplings are exponentially enhanced with respect to this parameter)
as this would result in the Yukawa matrix
entries becoming non-perturbative. Although there are some data points that satisfy all our conditions with large ${\rm Im}[\theta]$ around $M_1\sim10^9$ GeV, we find that, in general, $|{\rm Im}[\theta] |\lesssim 2$ is preferred. Ignoring effects from quantum gravity, a large portion of the points, close to 75$\%$, that satisfy perturbativity and vacuum stability up to the Planck scale, also satisfy these constraints up to the Landau pole of the hypercharge $\sim 10^{40}$ GeV.


\subsection{Asymptotically Safe $\SU (3)_c$ and Leptogenesis}
\label{sec:safety}

In the SM, the $\beta$-function of the strong coupling is negative, and therefore, the coupling decreases at higher energies \cite{Gross:1973id,Politzer:1973fx}. This property, known as asymptotic freedom, implies that quantum chromodynamics (QCD) is well-defined at arbitrarily short distances. Despite this, the running of $g_3$ has only been measured up to the TeV scale \cite{Bethke:2015etp} and the addition of new coloured particles could lead to the loss of this property. However, the possibility arises in which the strong gauge coupling reaches an interacting ultraviolet fixed point, as proposed in \cite{Sannino:2015sel}. Motivated by recent progress in constructing extensions of the Standard Model where at least one of the couplings becomes asymptotically safe, in this Section, we consider an extension of the SM in which QCD becomes asymptotically safe, rather than asymptotically free, in the UV. This is an alternative scenario where QCD is also well-defined at arbitrarily short distances and its cosmological consequences remain relatively unexplored. For this reason  we study its implications on leptogenesis with a type-I seesaw mechanism.\\

Asymptotic safety is achieved when the couplings in a model reach an interacting ultraviolet fixed point. These fixed points $g^*_i$ correspond to the zeros of the $\beta$-function $\beta_i(g^*_i)=0$. Asymptotic safety was recently shown to exist for gauge-Yukawa theories in a perturbative manner \cite{Litim:2014uca} and  has attracted recent attention \cite{Esbensen:2015cjw,Molgaard:2016bqf,Pelaggi:2017wzr,Abel:2017ujy,Bond:2017lnq,Abel:2017rwl,Bond:2017suy,Bond:2017tbw}. Theorems for weakly interacting theories with asymptotic safety have been established in \cite{Bond:2016dvk,Bond:2018oco}. In \cite{Bond:2017wut}, the authors provide a prescription for constructing extensions of the SM in which the weak and strong coupling constants reach perturbative fixed points in the UV, but the hypercharge still suffers from a Landau pole.\\

An alternative approach to achieve an interacting UV-fixed point for a gauge coupling, associated to the gauge group $G$, is to add a large number ($N_F$) of fermions charged under $G$ and perform a $1/N_F$ expansion in the computation of the $\beta$-functions \cite{PalanquesMestre:1983zy,Gracey:1996he,Holdom:2010qs}\footnote{For a different proposal to achieve asymptotic safety due to an energy cut off in the theory above which there are no quantum fluctuations see \cite{Khoze:2017lft}.}. Recently, the large-$N_F$ resummed contributions to the RG equations were computed in \cite{Antipin:2018zdg} for generic semi-simple groups. In \cite{Mann:2017wzh}, a large number of vector-like fermions were added to the SM in order to ensure the  asymptotic safety of the gauge couplings; nevertheless, 
this calculation was completed without the inclusion of the large-$N_F$ resummation for the Yukawa and the Higgs quartic. \\

The large-$N_F$ resummation was performed for a Yukawa coupling in \cite{Kowalska:2017pkt,Pelaggi:2017abg}. In the latter work, the resummation was also computed for a scalar quartic coupling. These results were applied in \cite{Pelaggi:2017abg} to extensions of the SM by a large number $N_F$ of charged fermions in order to make the strong or the weak gauge coupling asymptotically safe in the UV. Nonetheless, in that study it was shown that when one makes the hypercharge coupling safe in the UV, the Higgs quartic is driven to large non-perturbative values. This is because the location of the pole in the resummed functions for the Yukawa and the scalar quartic has the same location as the one in the Abelian case. \\

We now proceed to study a model where we add a large number ($N_F$) of coloured fermions in order to make the $\SU(3)_c$ coupling asymptotically safe and study the validity of a type-I leptogenesis scenario up to the scale at which the hypercharge develops a Landau pole $\sim 10^{40}$ GeV. With $N_F$ flavours of vector-like coloured fermions, the UV fixed point for the strong gauge coupling in the SM is given by,
\begin{equation}\label{eq:Deltab3}
g_3^*= \frac{4\pi}{\sqrt{\Delta b_3}}, \hspace{10mm} {\rm and} \hspace{10mm}  \Delta b_3 = \frac{2}{3} N_F S_{R_3} D_{R_2}, 
\end{equation}
where $\Delta b_3$ is the one-loop coefficient for Dirac fermions in the representation $R_i$, dimension $D_{R_i}$ and Dynkin index $S_{R_i}$. The large-$N_F$ resummed $\beta$-function has a pole at $g_3^*\!=\!4\pi/\sqrt{\Delta b_3}$ where it diverges to negative infinity. Before reaching the pole, this resummed contribution cancels out the one-loop contribution and the $\beta$-function becomes zero. For a review on the large-$N_F$ expansion we refer the reader to Ref.~\cite{Holdom:2010qs}.\\

\begin{figure}[t]
\centering
\includegraphics[width=.485\textwidth]{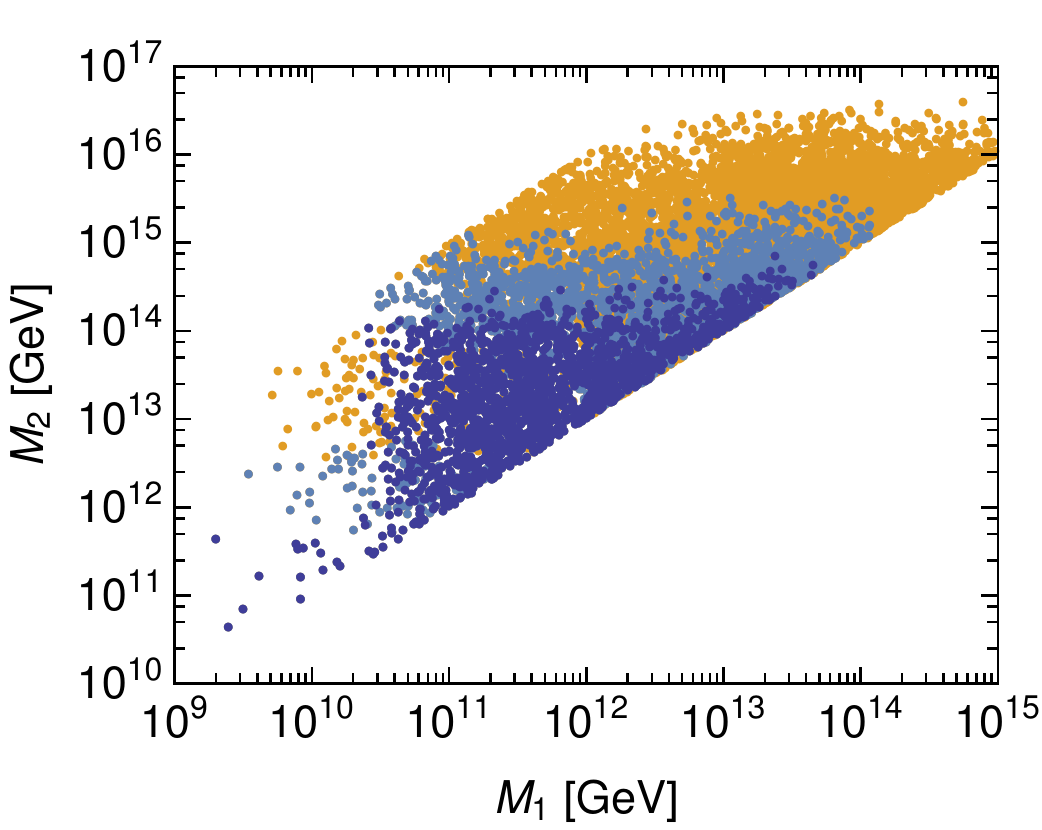} \,\,
\includegraphics[width=.485\textwidth]{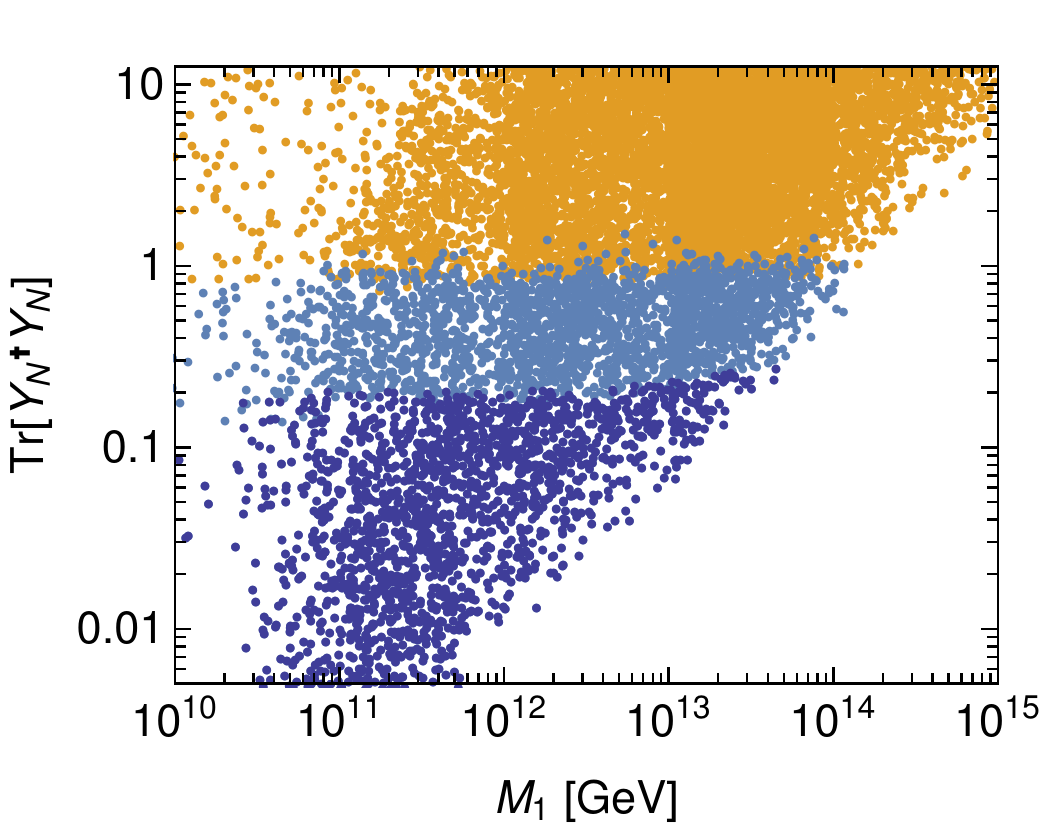}  \,\,
\includegraphics[width=.485\textwidth]{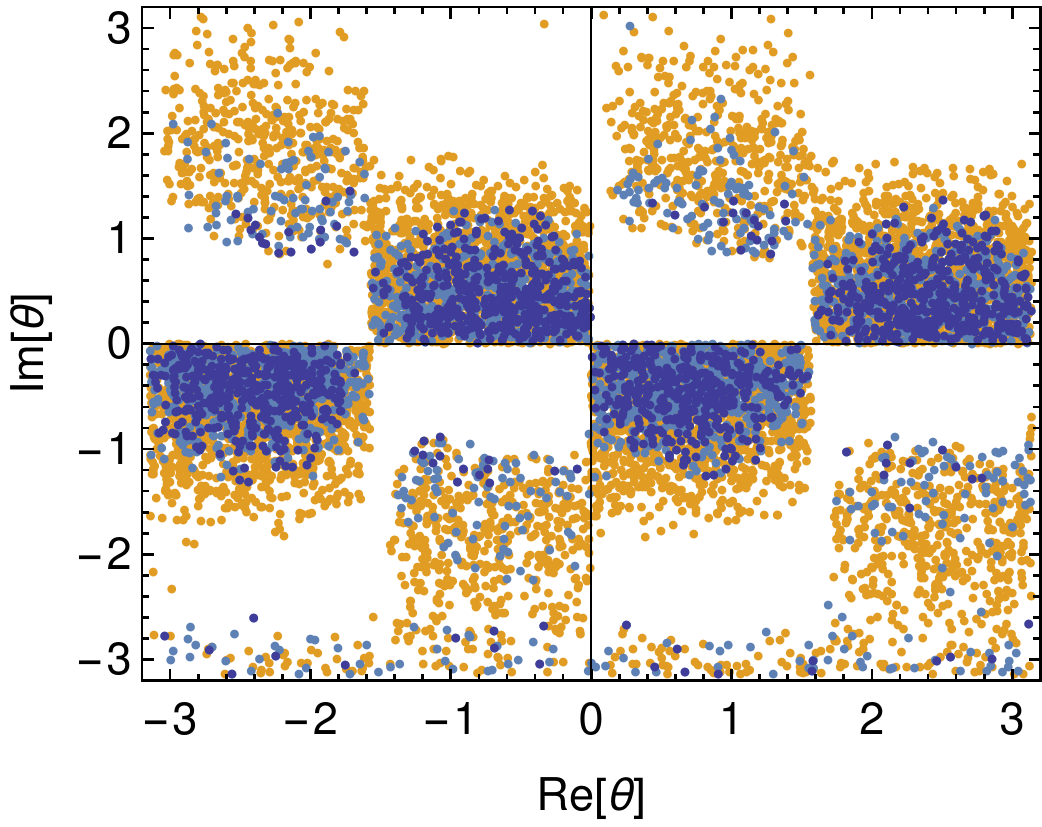}     \,\,
\includegraphics[width=.485\textwidth]{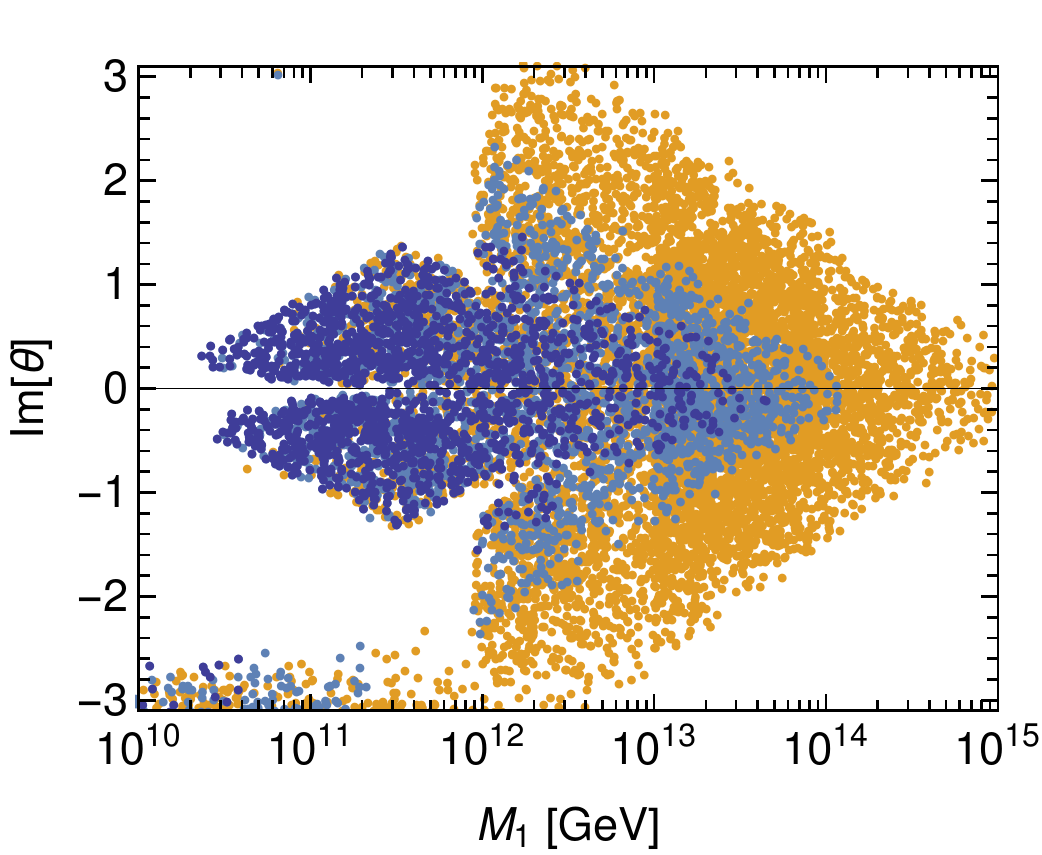}
\caption{Scatter plot for all the points that lead to successful leptogenesis with an asymptotically safe $\SU(3)_c$. Points in orange correspond to the points where any of the Lagrangian coupling constants become non-perturbative at some high energy scale and/or lead to a metastable EW vacuum with a lifetime smaller than the age of the Universe. Points in blue satisfy the perturbativity condition and the EW vacuum is metastable. For points in dark blue, the electroweak vacuum is absolutely stable. The RG study is performed up to $10^{40}$ GeV.}
\label{fig:scatter3}
\end{figure}

In order to perform the RG study, we consider the one-loop $\beta$-functions with the resummed contribution at leading order in $1/N_F$ which we present in \appref{sec:resummedRGEs}. For this scenario, we assume that gravity will only have subleading effects on the RG running of the couplings above the Planck scale, similar to the approach of softened gravity presented in Ref.~\cite{Giudice:2014tma}. In lack of a theory of quantum gravity, we will allow ourselves such speculation. \\

At one-loop the $N_F$ fermions give a positive contribution to the running of $g_3$ so that $g_3$ increases before reaching its fixed point, i.e. $g_3^*\geq g_3(M_t)$. Therefore, the negative contribution from $g_3$ to the running of $y_t$ can be larger, driving $y_t$ to smaller values compared to the SM case. This helps with the stability of the EW vacuum.
As was shown in Ref.~\cite{Pelaggi:2017abg}, the addition of $7\leq N_F \leq 13$ Majorana colour octets stabilises the EW vacuum. Here the lower bound derives from the requirement of convergence in the $1/N_F$ expansion for fermions in the adjoint representation \cite{Antipin:2017ebo}. The achievement of  vacuum stability provides  a further motivation to consider such a scenario. \\

For our numerical scan we consider a scenario with $N_F=13$ Majorana colour octets with common masses $M_F=5$ TeV. This corresponds to $\Delta b_3=26$ and $g_3^*\simeq 2.5$, for the one-loop coefficient and the fixed point for the strong coupling respectively. The calculation of the baryon asymmetry follows the same procedure from before with  the exception that the number of relativistic degrees of freedom is now $g_*\!=\!288.75$ due to the inclusion of the new vector-like fermions. 
The modification of $g_*$ must be accounted for in the density matrix equations; naturally, the Hubble rate is altered and this affects the decay parameter, $K_{1}$, (as detailed in \appref{sec:DMEapp}) which in turn 
affects the decay and washout of the lepton asymmetry. In addition, the Hubble-normalised thermal width of the tau and muon charged leptons are changed accordingly. The effect of an enlarged $g_*$ causes suppression of the decay parameter and the washout by a factor of a few. The corresponding results can be seen in Figs.~\ref{fig:scatter3}.\\

We would like to highlight the similarities and differences in the parameter space between the minimal type-I seesaw leptogenesis scenario and the large-$N_F$ case considered here. In the latter case, there exists a region in parameter space where the Higgs potential becomes absolutely stable. In addition, the bound on the neutrino Yukawa couplings is relaxed from ${\rm Tr}[Y_N^\dagger Y_N]\lesssim0.66$ (Fig.~\ref{fig:scatter1}) in the minimal case to ${\rm Tr}[Y_N^\dagger Y_N]\lesssim1.51$ in the large-$N_F$ case (Fig.~\ref{fig:scatter3}). The upper bound on $M_1$ is the same in both scenarios. \\

Finally, we note that the addition of a large number of fermions ($N_F$) that possess weak charge can make the $\SU(2)_L$ gauge coupling asymptotically safe. Nevertheless, in this scenario the Higgs quartic coupling violates the metastability bound \cite{Pelaggi:2017abg} and the addition of heavy Majorana neutrinos, which give a negative contribution to the the RG equation of $\lambda_H$, cannot improve this situation. Hence, we do not consider this scenario any further.


\section{Conclusions}
\label{sec:Conclusions}

The origin of neutrino masses and the baryon asymmetry of the Universe 
remain two experimentally observed phenomena that cannot be explained by the Standard Model. 
One of the most minimal solutions to these shortcomings is the introduction of  three heavy Majorana neutrinos to the SM particle
spectrum. As a consequence, the active neutrinos acquire their masses via a type-I seesaw mechanism and the baryon asymmetry is produced via CP-violating, out-of-equilibrium decays of the lightest sterile neutrino. \\

It is well known that in order to address the baryon asymmetry of the Universe using thermal leptogenesis with a hierarchical mass spectrum in the heavy neutrino sector requires $M_1 \gtrsim 10^{9}$ GeV. This means that the Yukawa couplings can be relatively large and hence the model may suffer from
non-perturbative couplings or violate the condition of metastability for the electroweak vacuum. In the present work we have discussed the self-consistency of thermal leptogenesis up to high scales. \\

We started by performing a numerical scan in order to find the points that produce sufficient baryon asymmetry, where each point in our scan  reproduces the central values for the measured neutrino mass differences (with normal ordering) and mixing angles. In order to compute the baryon asymmetry we have solved the density matrix equations that are valid in the whole range of masses studied. We subsequently solved the RG equations for these points in order to assess the validity of the model up to the Planck scale.\\

In summary, we have identified the region of parameter space in minimal thermal leptogenesis that allow for the \textbf{(i)} electroweak vacuum to remain metastable and \textbf{(ii)} for all the couplings in the model to remain perturbative up to the Planck scale. We find an upper bound on the neutrino Yukawa matrix and on the mass of the lightest right handed neutrino, ${\rm Tr}[Y_N^\dagger Y_N] \lesssim 0.66$ and $M_1\lesssim10^{14}$ GeV respectively. There is also a strong preference for $|{\rm Im}[\theta]| \lesssim 2$, where $\theta$ is the complex angle that parametrises the $R$ matrix. Detailed results can be seen in Figs.~\ref{fig:scatter1}-\ref{fig:scatter2}. Three quarters of the points that satisfy these constraints  up to the Planck scale also satisfy the constraints up to the scale at which the hypercharge develops a Landau pole $\sim 10^{40}$ GeV. Motivated by this and recent progress in making the gauge couplings in the SM asymptotically safe in the UV, we also studied a scenario where the $\SU(3)_c$ coupling becomes asymptotically safe and found constraints in the parameter space that leads to successful leptogenesis. However, this minimal model  still suffers from a Landau pole developed by the hypercharge coupling. It remains a future task to address this issue.

\acknowledgments

We would like to thank Bogdan Dobrescu, Francesco Sannino and Alessandro Strumia for helpful comments on the manuscript and Peizhi Du, Kristian Moffat, Cedric Weiland and Ye-Ling Zhou for useful discussions on thermal leptogenesis. We also thank Valentin Khoze and Luca Di Luzio for interesting discussions on the issue of vacuum stability. SI is supported by the University of California President's Postdoctoral Fellowship and partially by NSF Grant No.~PHY-1620638. ADP gratefully acknowledges financial support from CONACyT. This manuscript has been authored by Fermi Research Alliance, LLC under Contract No. DE-AC02-07CH11359 with the U.S. Department of Energy, Office of Science, Office of High Energy Physics.

\newpage
\appendix
\section{Details of Kinetic  Equations}\label{sec:DMEapp}
In this Appendix, we provide the formulae used within the density matrix equation \equaref{eq:full3}. The decay and washout terms are given by 
\begin{equation}
D_{1}(z) = K_{1} z \frac{\mathcal{K}_1 (z)}{\mathcal{K}_2 (z)}, \quad
\mathcal{W}_{1}(z) = \frac{1}{4} K_{1} \mathcal{K}_1 (z) z^3,
\end{equation}
where $\mathcal{K}_1$ and $\mathcal{K}_2$ are modified Bessel functions of the second kind with the decay asymmetry 
$K_{1}$ written as
\begin{equation}
K_{1}\equiv\frac{\tilde{\Gamma}_{i}}{H (T=M_1)}, \quad \tilde{\Gamma}_1= \frac{M_1 \left(Y^{\dagger} Y\right)_{11}}{8 \pi}.
\end{equation}

These CP-asymmetry parameters may be written as \cite{Covi:1996wh,Blanchet:2011xq,Abada:2006ea,DeSimone:2006nrs} 
\begin{equation}
\begin{aligned}
\epsilon^{(i)}_{\alpha\beta}=\frac{3}{32\pi\left(Y^{\dagger} Y\right)_{ii}}
\sum_{j\neq i}&\Bigg\{ i[Y_{\alpha i}Y^{*}_{\beta j}(Y^{\dagger}Y)_{ji}-Y_{\beta i}Y^{*}_{\alpha j}(Y^{\dagger}Y)_{ij}] f_1\left(\frac{x_{j}}{x_{i}}\right) \\
&+i[Y_{\alpha i}Y^{*}_{\beta j}(Y^{\dagger}Y)_{ij}-Y_{\beta i}Y^{*}_{\alpha j}(Y^{\dagger}Y)_{ji}] f_2\left(\frac{x_{j}}{x_{i}}\right) \Bigg\},
\label{eq:CPoff}
 \end{aligned}
\end{equation}
where
\begin{equation}\label{eq:CPa2}
f_1\left(\frac{x_{j}}{x_{i}}\right)=\frac{\xi\left(\frac{x_{j}}{x_{i}}\right)}{\sqrt{\frac{x_{j}}{x_{i}}}},\quad
 f_2\left(\frac{x_{j}}{x_{i}}\right)=\frac{2}{3\left(\frac{x_{j}}{x_{i}}-1\right)},\quad
 \xi\left(x\right) = \frac{2}{3}x\left[ \left(1+x\right)\ln\left( \frac{1+x}{x}  \right) -\frac{2-x}{1-x}   \right].
\end{equation}

The  density matrix equations of \equaref{eq:full3} may be expanded straightforwardly in terms of the flavour indices:
\begin{equation}\label{eq:longDM}
\begin{aligned}
\frac{dn_{N_{1}}}{dz}&=-D_{1}(n_{N_{1}}-n^\text{eq}_{N_{1}})\\
 \frac{dn_{\tau\tau}}{dz} &= \epsilon^{(1)}_{\tau \tau}D_{1} (n_{N_{1}}-n^\text{eq}_{N_{1}})   -\mathcal{W}_{1} \left\{
 \left| Y_{\tau 1}\right| {}^2   n_{\tau \tau } +  \text{Re}\left[Y^*_{\tau 1} \left( Y_{e1}n_{\tau e}+Y_{\mu 1} n_{\tau \mu}\right)\right] \right\}\\
 \frac{dn_{\mu\mu}}{dz} &= \epsilon^{(1)}_{\mu \mu}D_{1} (n_{N_{1}}-n^\text{eq}_{N_{1}}) -\mathcal{W}_{1} \left\{
\left| Y_{\mu 1}\right| {}^2   n_{\mu \mu } + \text{Re}\left[Y^*_{\mu 1}\left(  Y_{e1}n_{\mu e} + Y_{\tau 1}   n^*_{\tau \mu}  \right)\right] \right\} \\
 \frac{dn_{ee}}{dz} &= \epsilon^{(1)}_{ee}D_{1} (n_{N_{1}}-n^\text{eq}_{N_{1}}) -\mathcal{W}_{1} \left\{
 \left| Y_{e 1}\right| {}^2   n_{ee }  + \text{Re}\right[Y^*_{e1} \left(Y_{\mu 1}n^*_{\mu e} + Y_{\tau 1} n^*_{\tau e} \right )\left] \right\} \\
 \frac{dn_{\tau \mu}}{dz}& = \epsilon^{(1)}_{\tau \mu}D_{1} (n_{N_{1}}-n^\text{eq}_{N_{1}})-\left(\frac{\text{Im} \left( \Lambda_{\tau}  \right)}{Hz} + \frac{\text{Im} \left( \Lambda_{\mu}  \right)}{Hz}\right)n_{\tau \mu}  \\
 &-\frac{\mathcal{W}_{1}}{2} \left\{   n_{\tau \mu} \left( \left| Y_{\tau 1}\right| {}^2+ \left| Y_{\mu 1}\right| {}^2 \right)+ Y^*_{\mu 1}Y_{\tau 1} \left( n_{\tau \tau } + n_{\mu \mu}\right) + Y^*_{e1}Y_{\tau 1} n^*_{\mu e} + Y^*_{\mu 1}Y_{e1}n_{\tau e}  \right\}\\
 \frac{dn_{\tau e}}{dz} &= \epsilon^{(1)}_{\tau e}D_{1} (n_{N_{1}}-n^\text{eq}_{N_{1}}) -\frac{\text{Im} \left( \Lambda_{\tau}  \right)}{Hz}n_{\tau e}\\ 
 &-\frac{\mathcal{W}_{1}}{2} \left\{  n_{\tau e}\left( \left| Y_{e 1}\right| {}^2+ \left| Y_{\tau 1}\right| {}^2\right) +  Y^*_{e1}Y_{\tau 1}\left( n_{ee} + n_{\tau \tau }  \right)   + Y^*_{\mu 1}Y_{\tau 1} n_{\mu e} + Y^*_{e1}Y_{\mu 1}n_{\tau \mu}
\right\} \\
 \frac{dn_{\mu e}}{dz} &= \epsilon^{(1)}_{\mu e}D_{1} (n_{N_{1}}-n^\text{eq}_{N_{1}})-\frac{\text{Im} \left( \Lambda_{\mu}  \right)}{Hz}n_{\mu e}\\
 & -\frac{\mathcal{W}_{1}}{2} \left\{  n_{\mu e}\left( \left| Y_{e 1}\right| {}^2+ \left| Y_{\mu 1}\right| {}^2\right) + Y^*_{e1}Y_{\mu 1}\left( n_{ee} + n_{\mu \mu }  \right)   + Y^*_{e1}Y_{\tau 1}n^*_{\tau \mu} + Y^*_{\tau 1} Y_{\mu 1} n_{\tau e}
\right\},  
 \end{aligned} 
\end{equation}

where the total number density of the lepton asymmetry is given by the trace of this matrix, $n_{B-L}=n_{ee}+n_{\mu\mu}+n_{\tau\tau}$. In order to compare to measurement
of the baryon-to-photon ratio, $\eta_{B}$,  from cosmic microwave background measurements \cite{Ade:2015xua}, we have to ensure  $n_{B-L}$  is properly rescaled by a factor of  0.0096 which accounts for the sphaleron conversion factor (28/79) and the number density  of comoving photons after recombination \cite{Buchmuller:2004nz}.
The off-diagonal number densities encode the \emph{quantum correlations} between flavours differing flavours of charged leptons within the thermal bath; $n_{\alpha\beta}$ term describes the number density of $\ell_{\alpha}$ being converted to $\overline{\ell_{\beta}}$. \\

$\Lambda_{\alpha}$ represents the  self energy of the $\alpha$-flavoured  charged lepton in the thermal bath. The imaginary (real) part of the self energy is  the thermal width (mass) of the lepton and will be important in describing the transition between various flavour regimes of leptogenesis. Physically, the thermal width is related to the mean free path of the lepton in the thermal bath.\\

At high temperatures, ($T>10^{12}$ GeV), the Yukawa interactions of all the charged leptons are out of equilibrium and hence there is only one effective flavour of charged lepton.
As the Universe expands and cools, the interaction rates mediated by the SM $\tau$ Yukawa coupling come into thermal equilibrium.
At such temperatures the Universe can now distinguish flavour $\tau$ from $\tau^{\perp}$, where $\tau^{\perp}$ is  a coherent superposition of $e$ and $\mu$ flavour. 
As such,  the $\tau$-state has broken the coherence of the $N_{1}$ decays.
$\text{Im}(\Lambda_{\tau})$ enters into the Boltzmann equations in the following way
\begin{equation}\label{LT1}
\frac{\text{Im}(\Lambda_{\tau})}{Hz}\quad\text{where}\quad H(z)\sim 1.66 \sqrt{g^*}\frac{M^2_{1}}{M_{\rm PL}}\frac{1}{z^2}\implies zH(z)=1.66 g^*\frac{M^2_{1}}{M_{\rm PL}}\frac{1}{z}
\end{equation}
where  $z=\frac{M_{1}}{T}$. The thermal width of the $\tau$ takes the following form \cite{Blanchet:2011xq}
\begin{equation}\label{LT2}
\text{Im}(\Lambda_{\tau})=8\times 10^{-3}f^2_{\tau}T.
\end{equation}
Using (\ref{LT1}) and (\ref{LT2}) we find,
\begin{equation}\label{Lamb}
\frac{\text{Im}(\Lambda_{\tau})}{Hz}=\frac{8\times 10^{-3} f^2_{\tau}M_{\rm PL}}{1.66 \sqrt{g^*}M_{1}},
\end{equation}
where $M_{\rm PL}\!=\!1.22\times 10^{19}$ GeV and $g^*\!=\!106.75$. Note that $f_{\tau}$ is the $\tau$ charged lepton Yukawa coupling
\begin{equation}
m_{\tau}=f_{\tau} v\implies f_{\tau}=\frac{m_{\tau}}{v}=\frac{1.777}{174}\sim0.01,
\end{equation}
where all the units above are in GeV. We can rewrite (\ref{Lamb}) 
\begin{equation}
\frac{\text{Im}(\Lambda_{\tau})}{Hz}=4.85\times 10^{-8}\frac{M_{\rm PL}}{M_{1}}.
\end{equation}
As $\text{Im}(\Lambda_{\alpha})/Hz \propto 1/M_1$, the lower the scale of leptogenesis the larger impact the Hubble normalised thermal width. From Eq.~\eqref{eq:full3}, it can be seen that $\text{Im}(\Lambda)$ acts in competition with the off-diagonal number densities.
At lower values of $M_{1}$, the interaction rates of mediated by the SM charged lepton Yukawa couplings become increasingly significant. This means the $\tau$ and $\overline{\tau}$ in the thermal bath are more likely to find each other and annihilate via a Higgs than be  converted to an $e$ or $\mu$ via a heavy Majorana neutrino. This implies the flavour correlations between the flavours of charged leptons become less significant for smaller $M_{1}$.

\section{Resummed Contribution to the RGEs with a Large Number of Flavours}
\label{sec:resummedRGEs}

We consider an extension to the SM where we introduce a large number of coloured particles, this implies that only the running of $g_3$ and $y_t$ will be modified by the resummation at leading order in $1/N_F$. The resummed $\beta$-function for the strong gauge coupling can be written as,
\begin{equation}\label{eq:resummedalpha3}
\frac{\partial \alpha_3}{\partial \ln \mu} = -7 \frac{\alpha_3^2}{2\pi} + \Delta b_3 \frac{\alpha_3^2}{2\pi} + F_3 \left( A \right) \frac{\alpha_3^2}{2\pi},
\end{equation}
where $A\equiv\Delta b_3 \alpha_3/(4\pi)$, $\alpha_3 = g_3^2/(4\pi)$ and $\Delta b_3$ is the one-loop contribution from the $N_F$ fermions and is given in Eq.~\eqref{eq:Deltab3}. The resummed contribution is given by \cite{PalanquesMestre:1983zy, Gracey:1996he, Holdom:2010qs},
\begin{align}
F_3(A) & =  \int_0^{A} I_1(x) I_3(x) \, dx,\\
I_1(x) & \equiv \frac{(1+x)(2x-1)^2 (2x-3)^2 \sin^3(\pi x) \Gamma(x-1)^2 \Gamma(-2x)}{\pi^3 (x-2)}, \\ 
I_3(x) & \equiv \frac{4}{3} +\frac{3(20- 43 x +32 x^2 -14 x^3 +4x^4)}{2(2x-1)(2x-3)(1-x^2)}.
\end{align}
The $\beta$-function for the top Yukawa coupling is modified as follows,
\begin{equation}\label{eq:resummedyt}
(4\pi)^2 \frac{d y_t }{d \ln \mu} = \frac92 y_t^3 - y_t \left( 8 g_3^2 R_y(A_3)+ \frac94 g_2^2  + \frac{17}{12} g_Y^2 \right),
\end{equation}
where the resummed function is given by,
\begin{equation}\label{eq:Ry}
R_y(A) = \frac{2(3-2 A)^2(2-A) \sin (\pi  A) \Gamma (2-2 A) }{9 \pi A   \Gamma (3-A)^2},
\end{equation}
and hence at the fixed point, where $\alpha_3^*=4\pi/\Delta b_3$, $R_y(1)=1/9$. For details on the class of diagrams that contribute to the resummation and a detailed derivation of the resummed function $R_y(A)$ we refer the reader to Refs.~\cite{Kowalska:2017pkt,Pelaggi:2017abg}. For the scenario considered in Section~\ref{sec:safety} we solve the one-loop RG equations \eqref{eq:betagauge}-\eqref{eq:betaYN} with the equations for $g_3$ and $y_t$ replaced by \eqref{eq:resummedalpha3} and \eqref{eq:resummedyt} respectively. 

\newpage
\bibliographystyle{JHEP}
\bibliography{ref}{}
\end{document}